\documentclass[epj,nopacs]{svjour}

\usepackage{amsmath}
\usepackage{amssymb}
\usepackage{graphicx}
\usepackage{subcaption}
\usepackage{bm}
\usepackage{url}
\usepackage{multirow}
\usepackage{textcomp}
\usepackage{makecell}
\usepackage{booktabs}
\usepackage{siunitx}
\DeclareSIUnit{\charge}{\milli\volt\nano\second}
\DeclareSIUnit{\MeV}{\mega\electronvolt}
\DeclareSIUnit{\days}{days}
\usepackage[version=3]{mhchem}
\usepackage[colorlinks=true,linkcolor=blue,citecolor=blue,urlcolor=blue,hyperfootnotes=false]{hyperref}
\usepackage{doi}

\newcommand{\GEANT}{{\textsc{Geant}}}
\renewcommand{\makeheadbox}{}
\renewcommand{\ackname}{Acknowledgements.}
\hypersetup{pdftitle={Investigating production of TeV-scale muons in extensive air shower at 2400 m underground},pdfauthor={JNE Collaboration}}

\titlerunning{Investigating production of TeV-scale muons in EAS at 2400 m underground}
\authorrunning{X.~Zhang et al.}

\begin{document}

\title{Investigating production of TeV-scale muons in extensive air shower at 2400 m underground}

\subtitle{JNE Collaboration}

\author{Xinshun Zhang\inst{1,2} \and
Shaomin Chen$^{1,2,\ddagger}$ \and
Wei Dou\inst{1,2} \and
Haoyang Fu\inst{1,2} \and
Guanghua Gong\inst{1,2} \and
Lei Guo\inst{1,2} \and
Ziyi Guo\inst{1,2} \and
XiangPan Ji\inst{4} \and
Jianmin Li\inst{1,2} \and
Jinjing Li$^{3,1,\ast}$ \and
Bo Liang\inst{1,2} \and
Ye Liang\inst{1,2} \and
Qian Liu\inst{5} \and
Wentai Luo\inst{1,2} \and
Ming Qi\inst{7} \and
Wenhui Shao\inst{1,2} \and
Haozhe Sun\inst{1,2} \and
Jian Tang\inst{6} \and
Yuyi Wang\inst{1,2} \and
Zhe Wang$^{1,2,\dagger}$ \and
Changxu Wei\inst{1,2} \and
Jun Weng\inst{1,2} \and
Yiyang Wu\inst{1,2} \and
Benda Xu\inst{1,2} \and
Chuang Xu\inst{1,2} \and
Tong Xu\inst{1,2} \and
Tao Xue\inst{1,2} \and
Haoyan Yang\inst{1,2} \and
Yuzi Yang\inst{1,2} \and
Aiqiang Zhang\inst{1,2} \and
Bin Zhang\inst{1,2} \and
Yang Zhang\inst{8} \and
Zhicai Zhang\inst{1,2} \and
Lin Zhao\inst{1,2} \and
Yangheng Zheng\inst{5}}

\institute{Department of Engineering Physics and Center for High Energy Physics, Tsinghua University, Beijing 100084, China \and
Key Laboratory of Particle and Radiation Imaging (Tsinghua University), Ministry of Education, Beijing, China \and
School of Physics and Electronics, Hunan University, Changsha 410082, China \and
School of Physics, Nankai University, Tianjin 300071, China \and
School of Physical Sciences, University of Chinese Academy of Sciences, Beijing 100049, China \and
School of Physics, Sun Yat-Sen University, Guangzhou 510275, China \and
School of Physics, Nanjing University, Nanjing 210093, China \and
Institute of Frontier and Interdisciplinary Science, Shandong University, Qingdao 266237, China}

\date{Received: 18 April 2026 / Accepted: 1 September 2026\\
Author's final version. Published in \emph{Eur.\ Phys.\ J.~C} \textbf{86}, 1296 (2026).\\
\url{https://doi.org/10.1140/epjc/s10052-026-16340-5}}

\abstract{
Deep underground experiments present a new avenue to probe the first interactions in extensive air showers or hadronic interactions in the extreme forward phase space. The China Jinping Underground Laboratory, characterized by a vertical rock overburden of 2,400~m, provides an exceptionally effective shield against cosmic muons with energies below 3~TeV. The surviving high-energy muons, produced in the first interactions of extensive air showers, open a unique observational window into primary cosmic rays from tens of TeV up to the PeV scale and beyond. This distinctive feature also enables detailed studies of charged hadron production in the earliest stages of shower development. Using 1,338.6 live days of data collected with a one-ton prototype detector for the Jinping Neutrino Experiment, we measured the underground muon flux originating from air showers. The results show discrepancies of about 40\% corresponding to significances of more than 2$\sigma$, relative to predictions from several leading hadronic interaction models. We interpret these findings from two complementary perspectives: (i) by adopting the expected cosmic-ray spectra, we constrain the modeling of the first hadronic interactions in air showers and provide novel insights into resolving the long-standing \textit{muon puzzle}; and (ii) by assuming specific hadronic interaction models, we infer the mass composition of cosmic rays, and our data favor a lighter component in the corresponding energy range. Our study demonstrates the potential of deep underground laboratories to provide new experimental insights into air shower physics and cosmic rays.
}

\maketitle


\makeatletter
\def\@makefnmark{$^{\@thefnmark}$}
\def\@thefnmark{*}\@footnotetext{lijinjing@hnu.edu.cn (corresponding author)}
\def\@thefnmark{\dagger}\@footnotetext{wangzhe-hep@tsinghua.edu.cn}
\def\@thefnmark{\ddagger}\@footnotetext{chenshaomin@tsinghua.edu.cn}
\makeatother

\section{Introduction}

Cosmic rays (CRs) are charged particles originating from space, spanning energies from $10^9$ to $10^{20}$~eV and consisting of a chemical composition ranging from protons and helium nuclei to heavier elements such as oxygen and iron. Their origins, acceleration mechanisms, and propagation through the interstellar medium remain incompletely understood, but can be probed through measurements of their energy spectrum, chemical composition, and arrival direction~\cite{ParticleDataGroup:2024cfk}. Space-based experiments can directly measure these properties for CRs up to energies of about 100~TeV~\cite{AMS:2021lxc,Alemanno:2021gpb,CALET:2023nif}. At higher energies, indirect ground-based experiments with large effective areas are necessary, as the limited exposure of satellite detectors results in insufficient statistics.

When CRs traverse the atmosphere, they interact with atmospheric nuclei and initiate extensive air showers (EAS). These cascades produce numerous secondary particles, including hadronic, electromagnetic, and muonic components~\cite{Gaisser:2016uoy}. Ground-based observatories exploit these secondary particles as key observables to infer the properties of primary CRs~\cite{PierreAuger:2024flk,LHAASO:2024knt,KASCADEGrande:2011kpw}. Such measurements rely heavily on simulations of EAS, particularly on hadronic interaction models that extend into energy and phase-space regions inaccessible to current terrestrial accelerators. Detailed Monte Carlo (MC) simulations and other numerical approaches, combined with these models, are employed to describe EAS development~\cite{Heck:1998vt,Fedynitch:2018cbl}. However, data from the Pierre Auger Observatory reveal a persistent muon excess in EAS at EeV-scale compared to predictions from leading hadronic models (referred to as the \textit{muon puzzle})~\cite{PierreAuger:2016nfk}. Further study suggests that discrepancies accumulated across successive lower-energy sub-interactions during EAS development account for the observed muon excess~\cite{PierreAuger:2021qsd}. Experimental insights into sub-PeV and PeV-scale hadronic interactions in EAS can therefore offer direct evidence for the origins of the \textit{muon puzzle}.

For underground detectors, the kilometer-scale rock overburden filters out the abundant low-energy particles from EAS, making them particularly well suited for investigating the surviving TeV-scale muons. These energetic muons originate primarily from the first interactions of CRs with energies ranging from tens of TeV to PeV. Consequently, they serve as a unique probe for this energy regime, where measurements, particularly of medium- and heavy-mass components, remain sparse and highly uncertain in both space- and ground-based experiments~\cite{VERITAS:2018gjd,GRAPES-3:2024mhy,Gorbunov:2018stf}. Constraining the properties of CRs in this domain is essential for understanding their origins. Moreover, since these TeV-scale muons retain a substantial fraction of the primary energy, they carry information from the very first interactions in EAS. Their intensity thus directly probes the spectra of charged hadrons produced in very forward phase-space interactions (typically pseudorapidity $>10$) at TeV-scale center-of-mass (CM) energies, which is a region neither fully accessible nor well constrained in accelerator-based experiments~\cite{Albrecht:2021cxw}. While Large Hadron Collider (LHC) experiments such as LHCf and TOTEM focus on neutral hadrons and protons in the forward region, and FASER and SND@LHC indirectly infer charged hadron production via TeV-scale neutrinos~\cite{LHCf:2020hjf,TOTEM:2013pio,FASER:2024ref,SNDLHC:2024qqb}, underground muon measurements provide new complementary insights and constraints. Because the forward $\pi$ and $K$ decays producing these muons also generate the atmospheric neutrino flux, such measurements are pertinent to atmospheric neutrino physics as well~\cite{Sun:2025tng}.

Several experiments have probed high-energy muon production in EAS, a topic distinct from the \textit{muon puzzle} observed at EeV scales in ground-based observatories. In surface arrays, the information regarding the first interactions, carried by the number of GeV muons, is smeared by low-energy interactions; however, it can still be inferred indirectly from physical fluctuations in muon multiplicity~\cite{Cazon:2018gww,PierreAuger:2021qsd}. The KM3NeT underwater detector, sensitive to sub-TeV to multi-TeV muons, has reported atmospheric muon intensities exceeding predictions by about 40\% (ORCA) and 50\% (ARCA)~\cite{KM3NeT:2024buf}. Similarly, ground-based spectrometers like BESS and L3 have observed comparable excesses at sub-TeV energies~\cite{Haino:2004nq,L3:2004sed,Fedynitch:2021ima}. These excesses point to discrepancies between experimental data and model predictions for CRs at the tens-of-TeV energy scale, where the relevant hadronic interactions occur at CM energies of about 100~GeV. At higher primary energies, IceCube measures showers from 2.5~PeV to 100~PeV using a surface array in coincidence with muon bundles above 500~GeV detected in-ice~\cite{IceCube:2025baz}. However, these muons account for only a small fraction of the primary CR energy and are typically produced during intermediate EAS stages, where information from the first interactions is obscured by subsequent low-energy processes, akin to the situation in ground-based arrays.

\begin{figure}
    \centering
    \includegraphics[width=0.95\linewidth]{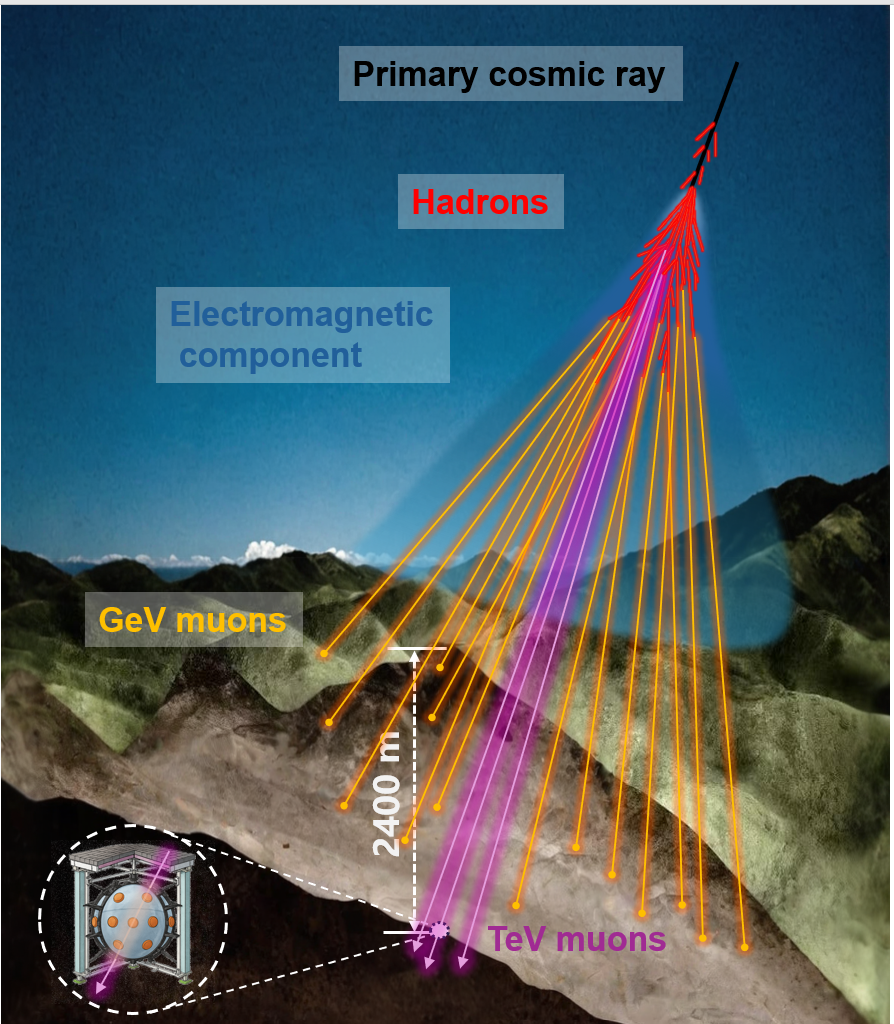}
    \caption{Schematic view of a TeV muon event observed with the one-ton prototype of JNE at CJPL.}
    \label{fig:mountain}
\end{figure}

The China Jinping Underground Laboratory (CJPL), characterized by a vertical rock overburden of 2,400 meters, offers an exceptional environment for studying muons above 3~TeV from EAS~\cite{Cheng:2017usi,JNE:2020bwn}. Since 2017, a one-ton prototype detector for the Jinping Neutrino Experiment (JNE), operating in CJPL-I (the first phase of CJPL), has measured the total muon flux with approximately 5\% precision~\cite{JNE:2024gov}. Fig.~\ref{fig:mountain} illustrates a schematic view of detecting a TeV muon from EAS using this detector. In this work, we present a precise measurement of the muon flux based on 1,338.6 live days of effective data collected by the one-ton prototype. We also perform detailed simulations of the underground muon flux using state-of-the-art CR and hadronic interaction models. By comparing the simulation results with measurements, this study makes a unique contribution to the understanding of CR properties and hadron production in the first interactions of EAS.

\section{The JNE detector at CJPL}

The main component of the one-ton prototype detector is a 20-mm-thick spherical acrylic vessel (1.29~m inner diameter) containing one ton of liquid scintillator (LS) as the detection target. When charged particles, such as muons, enter the target volume, their kinetic energies are deposited and numerous scintillation photons are produced for detection. This vessel is further enclosed by an outer cylindrical stainless steel tank (2.0~m diameter and 2.09~m height) filled with pure water. A total of 30 inward-facing 8-inch photomultiplier tubes (PMTs) are mounted on a stainless steel frame, positioned between the inner vessel and outer tank to detect scintillation photons. Each PMT is equipped with readout electronics capable of recording PMT waveforms at a resolution of 10 bits ADC over a 1~V dynamic range and a 1~GHz sampling rate for subsequent reconstruction and analysis. A 10-mm-thick black acrylic shield between the inner vessel and the PMTs blocks photons originating outside the target volume. Additionally, a 5-cm-thick lead layer is placed outside the detector, along with pure water in the outer tank, providing passive shielding against external radiation. A nitrogen bubbling system maintains positive internal pressure, effectively suppressing radon infiltration and improving scintillation photon yield by reducing oxygen quenching. Further details regarding this detector's design and functionality are referred to Refs.~\cite{Wang:2017ynm,Wu:2022oxo}.

\section{Event selection and reconstruction}

The detector collected data from July 31, 2017, to March 27, 2024. The data quality is monitored using electronics fluctuations and PMT occupancy, and about 5\% of the data are excluded. The deposited energy of collected events is primarily reconstructed by summing the total number of photoelectrons (PEs) on PMTs and further calibrated using intrinsic signals in the detector. To ensure a pure muon sample, the reconstructed energy is required to be greater than 90~MeV (typically a 45~cm track length in the LS) to exclude low-energy backgrounds. Two characteristic variables extracted from PMTs' waveforms, the number of peaks and the ratio of maximum PE to total PEs, further exclude intrinsic noise events from the detector, including accidental electronic disturbances and spontaneous light emissions from PMTs, with negligible efficiency loss. Finally, we select a total of 547 muon events from data for subsequent direction reconstruction and flux analysis. More details are introduced in Refs.~\cite{JNE:2020bwn,JNE:2024gov}.

We employ a template-based approach for directional reconstruction of muons, where templates obtained from MC simulations with detector responses that most closely match observed data events are selected. The muon direction is determined by computing a weighted average of the selected templates. In addition to the first photon hit time information from PMTs, the total number of collected PEs by each PMT is also incorporated into the detector response modelling. The reconstruction performance is assessed using an MC-simulated sample. Consequently, this technique exhibits a bias of less than 0.1$^\circ$ in reconstructing both zenith and azimuth angles, with an intrinsic angular resolution of 4.5$^\circ$ and consistent detection capability across the 4$\pi$ solid angle range.

Typically, muons deposit around 200~MeV of energy in the detector, usually leading to significant waveform saturation in the electronic readouts. To mitigate this issue, we developed a de-saturation algorithm that fits the rising and falling edges of saturated waveforms using linear and exponential functions, respectively, and extrapolates these fits to estimate the missing segments. The total number of PEs after de-saturation is taken as the photon intensity detected by the PMTs and used to correct the energy reconstruction. The deposited energy distribution of muon events after de-saturation closely matches the simulated distribution from MC simulations.

\section{Simulation framework}

We have developed a \GEANT4-based simulation framework to study muon detection and enable precise predictions~\cite{GEANT4:2002zbu,allison2006geant4}. The framework consists of three parts: muon production in EAS, propagation through the mountain, and detection in the one-ton prototype detector. We use a numerical method to obtain the differential energy and zenith angle distribution $\phi_s(E,\theta)$ of muon flux at sea level, while assuming an isotropic azimuthal distribution for high-energy muons ($>100~\mathrm{GeV}$)~\cite{Gaisser:2002jj}. We employ the Matrix Cascade Equations (MCEq) package~\cite{Fedynitch:2015zma,Fedynitch:2018cbl} to numerically solve the coupled cascade equations that describe the EAS development. This method requires inputs of the primary CR energy spectra of all chemical components, as well as a hadronic interaction model. Because the relevant hadronic interactions in EAS involve low momentum transfer, they occur in the non-perturbative regime of quantum chromodynamics (QCD), necessitating the use of effective theories and phenomenological models~\cite{Albrecht:2021cxw}. Therefore, we employ three phenomenological interaction models, which are tuned to LHC data. Among these models, SIBYLL-2.3d and QGSJET-II-04 were developed explicitly for EAS simulations, whereas EPOS-LHC was primarily designed for collider experiments but can also be applied to EAS simulations~\cite{Riehn:2019jet,Pierog:2013ria,Ostapchenko:2010vb}. For primary CR energy spectra, we use the state-of-the-art Global Spline Fit (GSF) model, which is derived from a global fit to experimental measurements and relies weakly on theoretical assumptions~\cite{Dembinski:2017zsh}.

\begin{figure}
    \centering
    \includegraphics[width=0.95\linewidth]{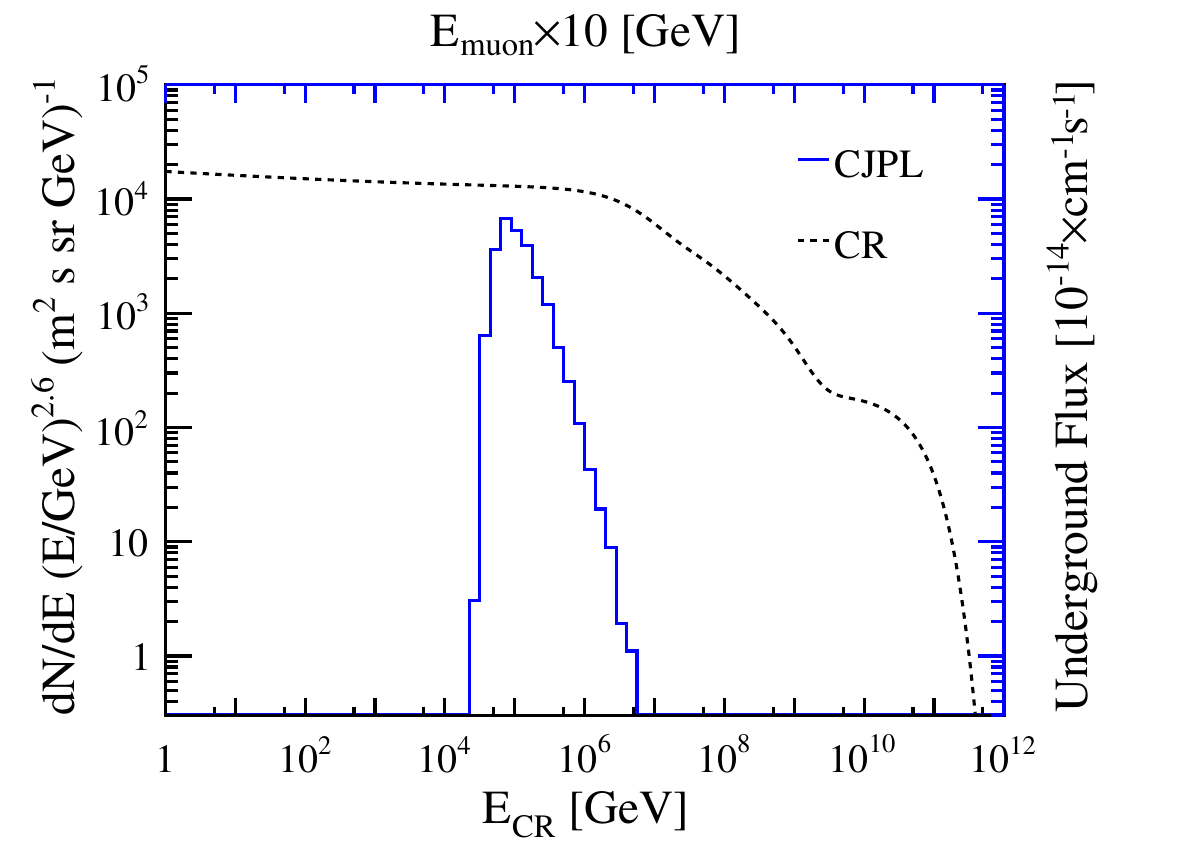}
    \caption{The surface energy distributions for muons arriving at CJPL-I, along with the CR total particle energy spectrum, as parameterized in Ref.~\cite{Gaisser:2011klf}. The hadronic model SIBYLL-2.3d is used here, accounting for EAS interactions. For comparison with primary CRs, the muon energy is multiplied by a factor, as discussed in Ref.~\cite{Fedynitch:2018cbl}.}
    \label{fig:surface}
\end{figure}

We utilized the terrain and rock models (marble with a density of 2.8~g/cm$^3$) surrounding CJPL to simulate the process of muons penetrating the mountain~\cite{CDEX:2021cll,zheng2024three}. The energy loss (ionization, radiation, and hadron interaction) of muons in rocks was simulated using \GEANT4, and surface events were sampled based on the muon distribution at sea level. As shown in Fig.~\ref{fig:surface}, the simulation indicates that muons entering CJPL require a surface energy exceeding 3~TeV (the simulation threshold is set at 2~TeV). Based on the calculations in Ref.~\cite{Fedynitch:2018cbl}, the corresponding primary CR energy is estimated to span from tens of TeV to PeV and above. Finally, the muon survival probability $P(E,\theta,\varphi)$ was calculated (MC error less than 0.5\%), and the low-probability cases with $\cos\theta < 0.2$ were excluded. The angular distribution of the underground muon flux $\phi_u(\theta,\varphi)$ is obtained by convolving the surface fluxes $\phi_s(E,\theta)$ and $P(E,\theta,\varphi)$. Then, muon events resulting from interactions within the fully configured detector, including all components such as electronics, are simulated and reconstructed using the same procedures as those applied in experimental data, as detailed in Refs.~\cite{JNE:2020bwn,JNE:2024gov}.

\section{Systematic uncertainties}

Simulations using post-LHC hadronic interaction models can predict the underground muon flux at CJPL-I. The total predicted flux is obtained by integrating the flux distribution $\phi_u(\theta,\varphi)$ over all solid angles. Fluxes within specific zenith angle ranges are calculated by integrating $\phi_u(\theta,\varphi)$ over those zenith intervals and all azimuthal angles. Major sources of systematic uncertainty include CR model, seasonal variations in muon flux, and uncertainties in the detector's position within the mountain. The uncertainties arising from hadronic interaction models are estimated by calculating the maximum variations of underground fluxes predicted by three hadronic models, while keeping the CR model fixed at the GSF model. Table~\ref{tab:uncertainty} summarizes the uncertainties affecting flux ratios between data and predictions; further details on systematics are discussed below.

\begin{table}[!htbp]
    \tabcolsep=0.3cm
    \centering
    \caption{The uncertainties in the measured and predicted muon flux. Statistical fluctuations in the prediction have been reduced to the 0.1\% level, which is negligible in this study.}
    \label{tab:uncertainty}
    \begin{tabular}{clc}
        \hline\noalign{\smallskip}
        &  & Uncertainty [\%] \\
        \noalign{\smallskip}\hline\noalign{\smallskip}
        \multirow{2}*{Measurement} & Statistics &  4.2 \\
        & Systematics & 2.2 \\
        \noalign{\smallskip}\hline\noalign{\smallskip}
        \multirow{4}*{Prediction} & Seasonal variation & 0.5 \\
        & Detector position &  1.6 \\
        & CR model & 11.8\\
        & Hadronic model & 4.8\\
        \noalign{\smallskip}\hline
    \end{tabular}
\end{table}

Seasonal changes in atmospheric temperature affect atmospheric density and, consequently, the total underground muon flux~\cite{Gaisser:2016uoy}. Local atmospheric temperatures around CJPL were obtained from the European Centre for Medium-Range Weather Forecasts~\cite{ERA52011}. Using the theoretical framework presented in Ref.~\cite{Grashorn:2009ey} and parameters from Ref.~\cite{MINOS:2009njg}, we calculated the conversion coefficient linking temperature fluctuations to muon flux variations at CJPL-I to be 0.95. We determined time-varying effective temperatures ($T_\mathrm{eff}$) by weighting temperatures across different atmospheric levels. We obtain the relative underground muon flux variations by multiplying the relative variations in $T_\mathrm{eff}$ by the conversion coefficient 0.95. The standard deviation of $T_\mathrm{eff}$ was approximately 1.1~K at different times, resulting in an estimated flux uncertainty of 0.5\%. Finally, we applied an uncertainty estimate that is independent of the zenith angle.

Determining the precise relative position of the one-ton prototype within the mountain is challenging due to the irregular terrain, which causes variations in the predicted underground muon flux depending on the assumed simulation positions. To address this uncertainty, we use a data-driven approach based on the sensitivity of the detected muon angular distribution to the detector position. By scanning positions around the nominal value in the simulation, we quantify the angular difference between observed data and simulations using a Pearson $\chi^2$ test. The minimum $\chi^2$ corresponds to the best fit point of detector position inside the mountain. We estimate the uncertainty by shifting the position until $\chi^2$ increases by one, resulting in an uncertainty of $\pm30$~m. It corresponds to a 1.6\% systematic uncertainty in the predicted muon flux. Furthermore, the angular dependence of this uncertainty exhibits significant variations, especially at large zenith angles, mainly owing to the non-uniform mountain terrain. In the flux measurement, positional uncertainty also propagates into the estimation of the effective area, introducing an additional 1.1\% uncertainty in the measured flux. Given that the detector position uncertainty affects both the measurement and prediction correlatively, we account for it only once in the comparative analysis to accurately assess its impact on the measurement-to-prediction ratio without double-counting.

In the simulations, the mountain is modeled as a uniform rock structure. Potential deviations, such as cavities or variations in density, are assessed using radiography to evaluate their impact on the underground muon flux $\phi_u(\theta,\varphi)$. The angular phase space is divided into $6 \times 6$ bins for $\cos\theta$ and $\varphi$, and muon attenuation in each bin is calculated by comparing surface flux predictions derived from MCEq with experimental data. Simulated attenuation profiles, based on penetration depth, facilitate the reconstruction of the average mountain depth in each direction, allowing for comparison with the nominal geometry. No angular bin shows deviations greater than $2\sigma$, which confirms the assumption of uniform rock density. In the previous measurement, an exaggerated elevation variation was intentionally chosen to explicitly demonstrate its negligible impact on detection efficiency and effective area~\cite{JNE:2024gov}. After calibrating the CJPL depth measurements against a 1560-meter reference platform, the uncertainty in elevation becomes negligible for flux predictions. Using the simulation framework, we also predict the underground flux at CJPL-II (the second phase of CJPL). The ratio between predicted fluxes of CJPL-I and CJPL-II, considering the kilometer-scale relative distance and mountain structure, is consistent with the comparison using measurements in Refs.~\cite{JNE:2024gov,zhang2025measurement}, thus corroborating the detector position and mountain model in this study.

The CR energy spectrum and mass composition in this study are derived from the data-driven GSF model, minimizing theoretical assumptions~\cite{Dembinski:2017zsh}. We also evaluate predicted fluxes using alternative models, including the H3a model~\cite{Gaisser:2011klf}, the Zatsepin-Sokolskaya model calibrated with PAMELA data (ZSP)~\cite{Zatsepin:2006ci,PAMELA:2011mvy}, and others, while keeping the hadronic interaction model unchanged at SIBYLL-2.3d. These models demonstrate varying CR total fluxes and mass compositions. In comparison to the GSF model, predictions from both the H3a and ZSP models exhibit significant deviations of $+11.7\%$ and $-11.8\%$, respectively. In this analysis, these deviations are considered as uncertainties originating from the CR models, which is comparable to evaluations presented in Ref.~\cite{KM3NeT:2024buf}.

In the measurement, the selection efficiency of muon events is evaluated using the generated MC sample by applying the same selection criteria as those in data. The muon events in simulation are generated on the surface of a rectangular box that encompasses the entire detector geometry to evaluate its effective area. Consequently, the selection efficiency for muon signals is estimated to be 82\% across all zenith angles, with clipping muons being the primary contributor to the efficiency loss. The effective area of the detector, which accounts for underground muon spectra, is calculated to be 1.58~m$^2$. Corresponding uncertainties are determined by varying the simulation setup used for generating the muon sample. The primary uncertainty regarding selection efficiency stems from detector geometry, followed by uncertainties related to energy scale. Geometry-related uncertainties directly influence simulated detector responses and thus affect efficiency estimations; these uncertainties have been evaluated at a total of 1.7\% through variations in detector geometry within simulations. Previously, encountered waveform saturation required a conservative extrapolation of muon energy-scale uncertainties from low-energy data; however, advancements in energy reconstruction and calibration have reduced this uncertainty to 1.1\%, resulting in a flux uncertainty of 0.5\%. Additionally, the effective-area uncertainty is assessed at 1.3\% by varying detector positions and surface muon distributions while evaluating their impact on underground muon spectra. As a result, the total muon flux at CJPL-I is $(3.54 \pm 0.15_{\mathrm{stat.}} \pm 0.08_{\mathrm{syst.}}) \times 10^{-10}~\mathrm{cm}^{-2}\mathrm{s^{-1}}$. Further details are given in Ref.~\cite{JNE:2024gov}.

To estimate the possible flux bias from unresolved multi-muon bundles, we simulate air showers with CORSIKA, weight them to the primary-energy distribution used for the CJPL muon sample, and propagate the muons through the mountain with \GEANT4, including multiple Coulomb scattering (MCS)~\cite{Heck:1998vt,GEANT4:2002zbu,allison2006geant4}. The calculation includes the detector geometrical acceptance and the simulated higher-multiplicity tail. Under conservative assumptions, it gives an upper estimate of approximately 0.08\% for the associated flux bias relative to simulated single-muon showers. This is far below the current measurement uncertainty.

In addition to TeV-scale muons produced directly in EAS, atmospheric neutrinos generated in these cascades can also yield secondary muons underground. These neutrinos, typically with GeV-scale energies, may traverse the mountain overburden or propagate through the Earth and subsequently undergo charged-current interactions in the surrounding rock. The resulting muons can then enter the detector volume. To quantify this contribution, we scale the neutrino-induced muon rates measured by Super-Kamiokande and SNO~\cite{Super-Kamiokande:2023ahc,SNO:2009oor}. For the one-ton prototype, the expected yield is 0.3 events per year, corresponding to less than 1\% of the total muon flux. This background is therefore negligible for the present study.

\section{Results}

\begin{figure}
    \centering
    \includegraphics[width=0.95\linewidth]{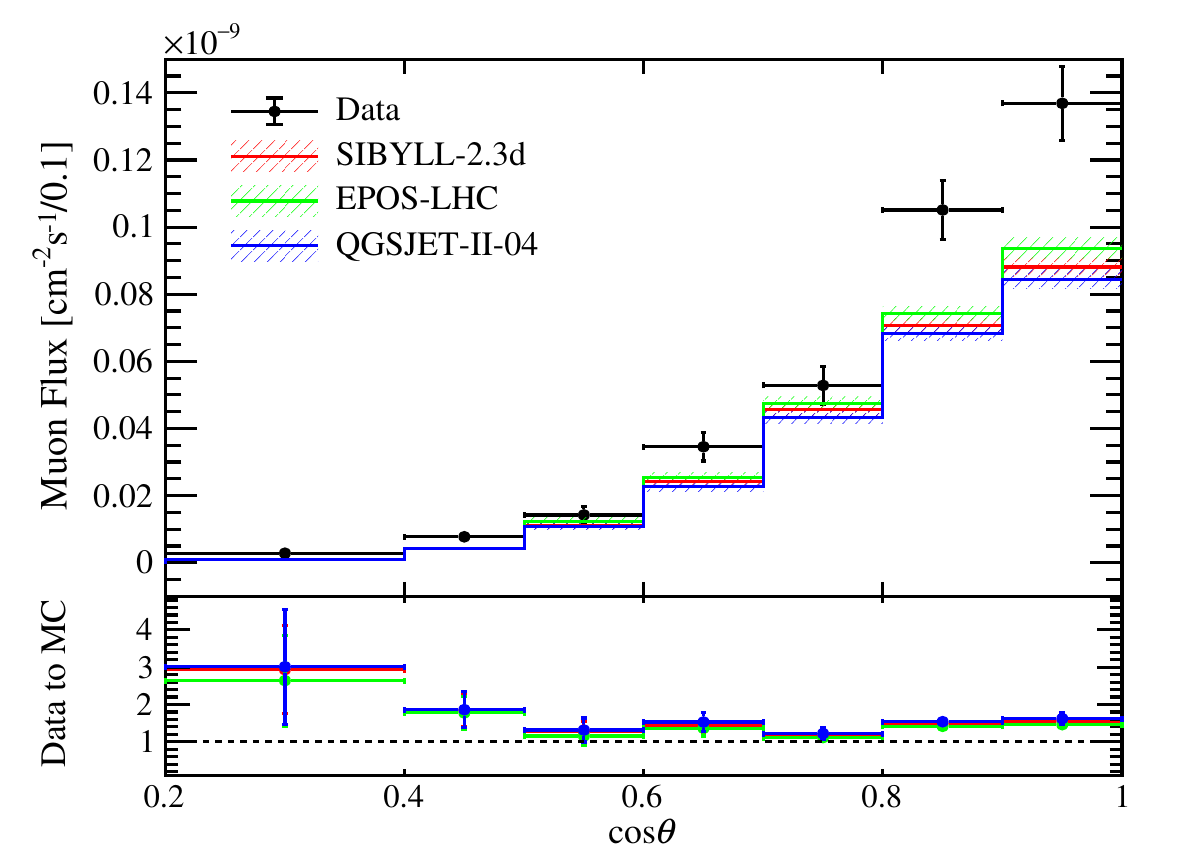}
    \caption{The comparisons of underground muon flux at CJPL-I between data and predictions based on post-LHC models at different zenith angles. In the top panel, the uncertainties from measurement points are plotted with error bars, while the uncertainties from predictions are shown with the dashed bands and model-related uncertainties are not included here. In the bottom panel, both measurement and prediction uncertainties are combined and plotted with error bars.}
    \label{fig:underground}
\end{figure}

Table~\ref{tab:flux} shows the ratios of muon flux between data and predictions, revealing a discrepancy of about 40\% with more than 2$\sigma$ (including model-related uncertainties) and 5.5$\sigma$ (excluding them). Thanks to the excellent angular resolution, we can systematically compare results based on the zenith angle, allowing us to investigate any potential dependence of flux excess on this angle. Muon fluxes in various zenith angle bins are calculated from reconstructed muon directions using the formula $\phi(\theta) = N(\theta)/\left[\epsilon(\theta)S(\theta)T\right]$, where $N(\theta)$ represents the number of muons, $T$ denotes the effective DAQ time, $\epsilon(\theta)$ and $S(\theta)$ correspond to the detection efficiency and effective area, respectively. We divide $\cos\theta$ values from 0.2 to 1 into bins of width 0.1 due to angular resolution and the statistical limitations; notably, the interval between 0.2 and 0.4 is combined into a single bin because of limited statistics. The average angular deviation caused by multiple Coulomb scattering for muons above 3~TeV is estimated to be 0.28 degrees, which is negligible compared to our angular resolution. The effective energy threshold for muons is determined by the slant depth of the mountain overburden, which varies with the zenith angle. For the angular range in this analysis, the threshold spans from 3~TeV to 9~TeV. Fig.~\ref{fig:underground} displays measured and predicted underground muon fluxes at different zenith angles, showing no significant angular dependence for the muon excess.

\begin{table}[!htbp]
    \tabcolsep=1.32cm
    \centering
    \caption{The muon flux ratios between data and simulation using different hadronic interaction models. Model-related uncertainties are not included here. These comparisons evidence about 40\% excess for underground muon flux.}\label{tab:flux}
    \begin{tabular}{cc}
        \hline\noalign{\smallskip}
        Model & Flux ratio  \\
        \noalign{\smallskip}\hline\noalign{\smallskip}
         SIBYLL-2.3d &  1.44$\pm0.07$      \\
         EPOS-LHC &  1.38$\pm0.07$     \\
         QGSJET-II-04 &  1.51$\pm0.08$    \\
        \noalign{\smallskip}\hline
    \end{tabular}
\end{table}

\section{Discussion}

\begin{figure}
    \centering
    \begin{subfigure}[b]{0.85\linewidth}
        \includegraphics[width=\linewidth]{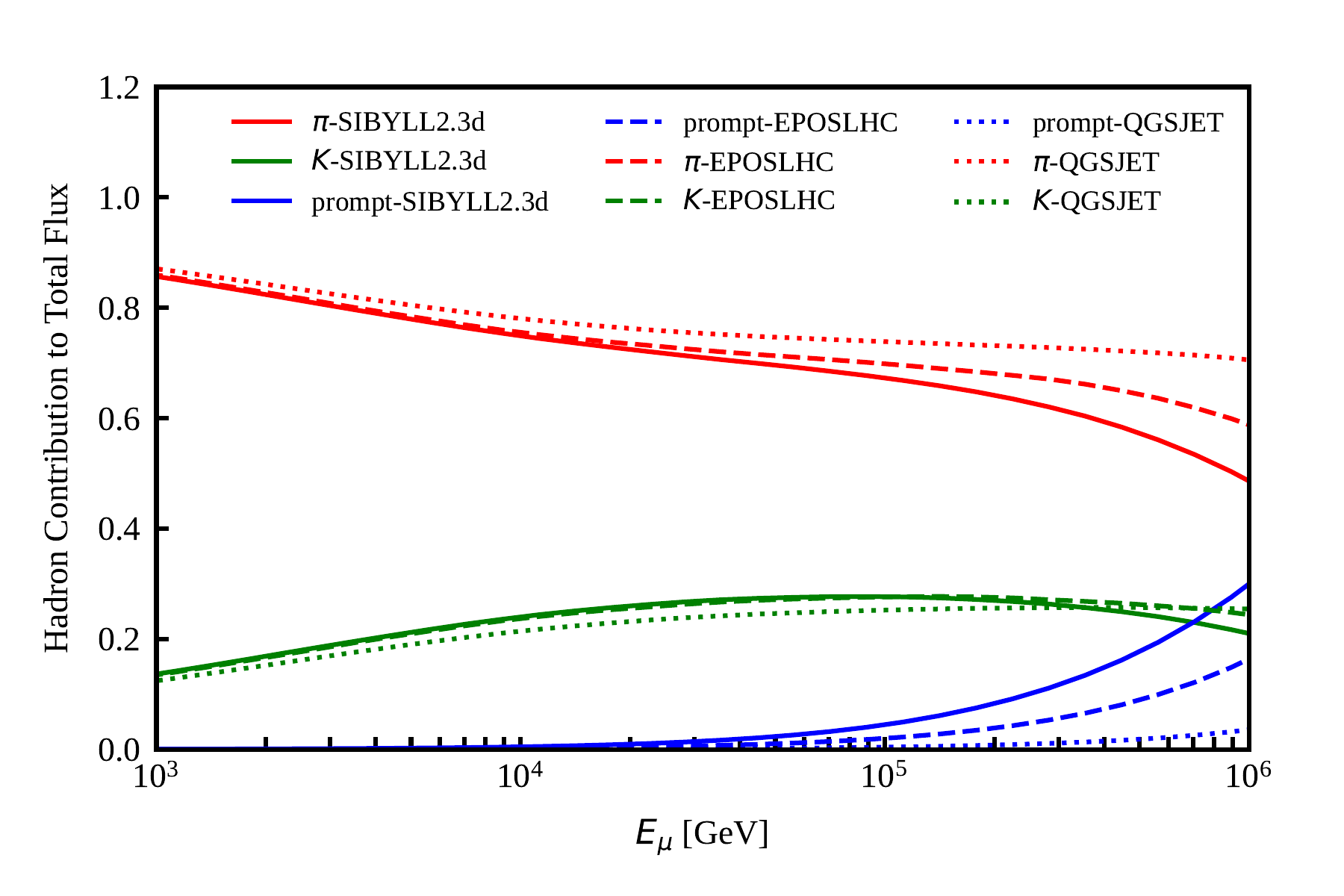}
        \caption{Fraction to Total Flux}
        \label{fig:fraction}
    \end{subfigure}

    \begin{subfigure}[b]{0.85\linewidth}
        \includegraphics[width=\linewidth]{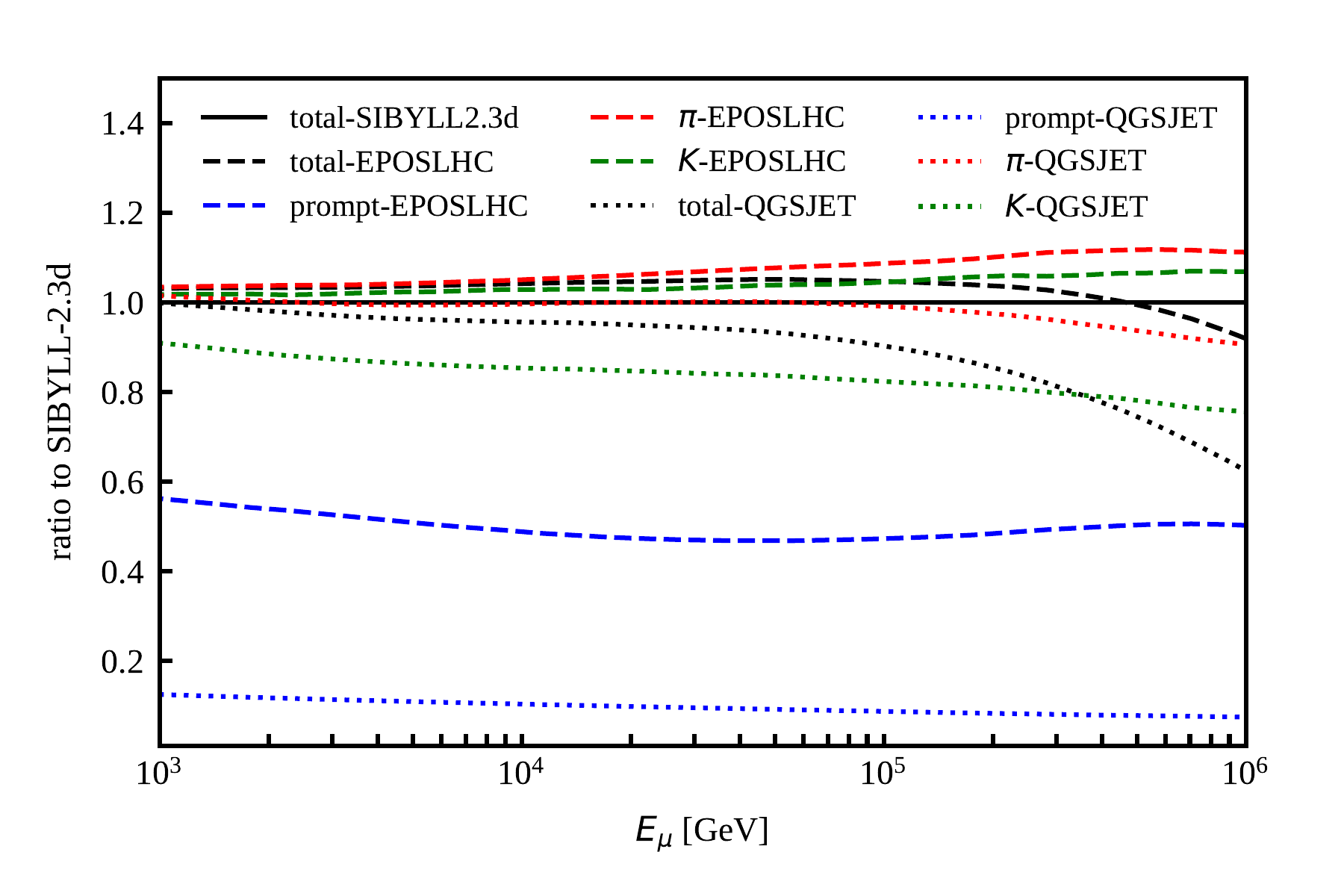}
        \caption{Comparison between Models}
        \label{fig:compare}
    \end{subfigure}
    \caption{Muon contributions from $\pi$, $K$, and prompt decays calculated with MCEq, using the GSF cosmic-ray model and integrated over zenith angle. Panel (a) shows each contribution as a fraction of the total muon flux; panel (b) shows ratios of the differential energy spectra to the corresponding SIBYLL-2.3d predictions, with the black lines denoting the ratio for the total flux. In both panels the red, green, and blue lines denote the $\pi$, $K$, and prompt contributions, and the solid, dashed, and dotted lines denote the SIBYLL-2.3d, EPOS-LHC, and QGSJET-II-04 models, respectively.}
    \label{fig:model}
\end{figure}

To understand the origin of the observed discrepancy from the standpoint of hadronic interactions in EAS, we compare the muon production mechanisms predicted by different hadronic interaction models, as shown in Fig.~\ref{fig:model}. Using MCEq, the calculated muon flux is decomposed into contributions from $\pi$ and $K$ mesons, as well as prompt sources arising from charm and unflavored mesons~\cite{Fedynitch:2018cbl}. Focusing on muons above the TeV scale highlights differences in meson production during the earliest stages of the cascade. The three models considered here exhibit distinct features: EPOS-LHC predicts enhanced production of both $\pi$ and $K$ relative to the other models, while QGSJET-II-04 yields comparable $\pi$ but a substantially lower $K$ component than SIBYLL-2.3d. At higher energies, the prompt component becomes increasingly important, with SIBYLL-2.3d producing a significantly larger fraction of such mesons.

The variations among the three hadronic models suggest potential phenomenological explanations for the observed discrepancy. A straightforward mechanism is a hardening of the charged hadron spectrum (dominated by $\pi$ production) in the first interactions of the cascade, which also reduces the hadron multiplicity and channels more energy into fewer leading mesons, thereby enhancing the yield of high-energy muons from their subsequent decays. However, these parameters are only weakly constrained by ground-based EAS measurements and accelerator-based experiments~\cite{Albrecht:2021cxw}. In the simplified Heitler--Matthews framework, the characteristics of the first interaction in EAS have minimal impact on primary observables in ground-based arrays, such as the GeV-muon content and the depth of shower maximum; however, the physical fluctuations in GeV-muon content across different showers can be utilized to infer the hadron spectra~\cite{Cazon:2018gww,Matthews:2005sd}. While studies of muon-content fluctuations for CRs in the 10~TeV to PeV range are limited, there remains considerable flexibility for tuning within hadronic interaction models. Consequently, the discrepancy reported herein may indicate a harder spectrum for charged hadrons and a smaller hadron multiplicity in the first interaction of EAS.

Alternatively, an enhancement in charged $K$ or prompt meson production can also increase the yield of high-energy muons. Charged kaons are more likely than charged pions of the same energy to decay before interacting in the atmosphere. Prompt mesons, such as $D^{\pm}$ and $D^{0}$, have much shorter lifetimes and generally decay before interacting at TeV energies. Yet the prompt contribution at the TeV scale is estimated at the percent level, as shown in Fig.~\ref{fig:fraction}, which is insufficient to account for the observed discrepancy~\cite{Fedynitch:2018cbl}. Similarly, the flux contribution of $K$ is about 25\%, and a 50\% increase of the $K$ yield produces only about 10\% more muons. Therefore, these estimates suggest that the observed discrepancy could not be explained independently by an enhancement of charged $K$ production, with unchanged hadron multiplicity.

Notably, enhanced $K$ production would simultaneously affect both underground and ground-level muon observables. In shower development, increased $K$ production comes at the expense of $\pi$ yield, including the neutral component $\pi^0$. Because $\pi^0 \to \gamma\gamma$ immediately transfers energy to the electromagnetic cascade, a reduction in $\pi^0$ production retains a larger energy fraction in the hadronic channel, thereby boosting the GeV-scale muon yield at the ground~\cite{Albrecht:2021cxw,Manshanden:2022hgf}. Although the excess of GeV-muon content in ground-based arrays is observed to set in around tens of PeV, studies at the sub-PeV scale remain limited and measurements of GeV-muon content at the PeV scale carry large systematic uncertainties at the tens-of-percent level~\cite{Soldin:2021wyv,IceCube:2022tla}. Assuming an enhancement of charged $K$ production starting from sub-PeV-scale interactions, the GeV-muon content excess for CRs above tens of PeV can be explained~\cite{Riehn:2024prp}. Independent support for this scenario comes from LHC data: strangeness enhancement has been observed within the mid-rapidity region and partly extended to the forward region, while the latest hadronic models cannot reproduce this effect~\cite{ALICE:2016fzo,Anchordoqui:2019laz,LHCb:2023rpm}. Additionally, seasonal variations of atmospheric neutrinos at the TeV scale exhibit a 2--3$\sigma$ discrepancy compared with hadronic model predictions, which also suggests enhanced $K$ production in EAS~\cite{IceCube:2023qem}.

\begin{figure}
    \centering
    \includegraphics[width=\linewidth]{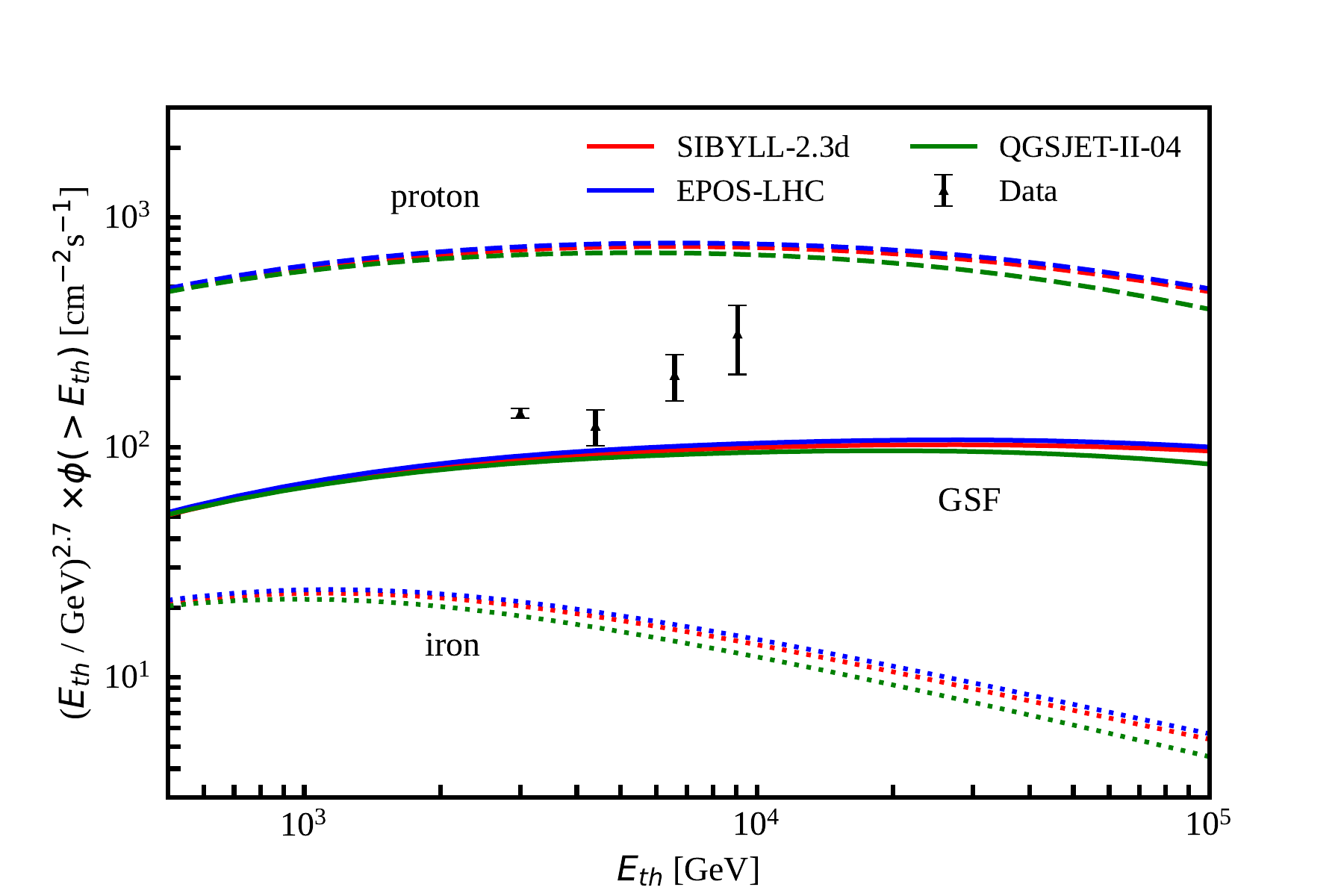}
    \caption{The integral energy spectra for muons with energies above the threshold $E_{th}$. The red, blue, and green lines represent the results calculated from SIBYLL-2.3d, EPOS-LHC, and QGSJET-II-04, respectively. The solid lines represent the GSF model, while the dashed and dotted lines are derived by assuming CRs with pure components of protons and iron. The experimental measurements are plotted with error bars. Details are explained in the text.}
    \label{fig:mass}
\end{figure}

The CR energy spectra of all mass components adopted in this work follow the data-driven GSF model, thereby minimizing theoretical assumptions~\cite{Dembinski:2017zsh}. This framework, however, relies critically on experimental measurements of individual CR components, which remain sparse or uncertain in the TeV--PeV range, particularly for medium- and heavy nuclei~\cite{VERITAS:2018gjd,GRAPES-3:2024mhy,Gorbunov:2018stf}. Our measurements thus provide new constraints on the CR composition in this energy window under different hadronic interaction models. In ground-based EAS experiments, the observed muons carry only a small fraction of the primary energy. Within the superposition picture, heavy primaries yield more total muons than light primaries. By contrast, the production of muon flux at $E_\mu$ is dominated by primaries with energies around $10E_\mu$~\cite{Fedynitch:2018cbl}, where light nuclei attain higher average nucleon energies than heavy ones~\cite{Miller:2007ri}. Consequently, light primaries have a greater probability of producing high-energy mesons capable of decaying into TeV-scale muons.

Using MCEq, we compute the integrated muon flux above the energy threshold $E_{th}$ for pure proton and pure iron primaries, adopting a broken power-law spectrum consistent with the CR knee~\cite{LHAASO:2024knt}. Results based on the GSF model are included for comparison. Experimentally, we extract the integrated muon fluxes with different energy thresholds by combining the measured underground fluxes under various $\theta$ with survival probabilities from mountain simulations, accounting for variations across hadronic models. For different $\theta$, the energy thresholds of penetrating muons are different due to variations in rock overburden depths. The muon events with $\cos\theta<0.6$ are divided into three groups as before, while those with $\cos\theta>0.6$ are combined due to similar energy thresholds. The measured fluxes, shown in Fig.~\ref{fig:mass}, systematically exceed the GSF predictions across all tested hadronic interaction models, favoring a lighter primary composition for CRs within 10~TeV--PeV range, with a significance greater than 5.5~$\sigma$ (excluding model-related systematics). The excess exhibits an upturn trend as the energy threshold increases, but is limited by the large statistical uncertainty.

Recent updates to major hadronic interaction models, specifically QGSJET-III, SIBYLL*, and EPOS-LHC-R, are primarily motivated by the long-standing \textit{muon puzzle}~\cite{Riehn:2024prp,Pierog:2025ixr,Ostapchenko:2024myl}. We subsequently performed EAS simulations using CORSIKA-7.8010~\cite{Heck:1998vt} for primary CRs with energies from tens of TeV to PeV to quantify the effect of these latest model revisions. The resulting enhancement in muon production above 3 TeV is marginal: $<8\%$ for EPOS-LHC-R, and $<5\%$ for both QGSJET-III and SIBYLL*~\cite{Riehn:2024prp}. Such minor enhancements do not alter the conclusions of this work. At the same time, increasingly precise measurements of CRs have become available. In particular, results from the Large High Altitude Air Shower Observatory indicate a lighter mass composition between 300~TeV and 30~PeV compared with earlier observations~\cite{LHAASO:2024knt}. Incorporating these data, the updated GSF framework potentially predicts a somewhat higher muon flux~\cite{Fujisue:2025wnp}, thereby reducing the significance of the discrepancy reported here.

\section{Summary}

In summary, we present the first study of muon production in air showers at CJPL, based on 1,338.6 live days of data collected with the one-ton prototype of the Jinping Neutrino Experiment. We report the first measurement of the differential muon flux at CJPL, with an angular reconstruction accuracy of approximately 4.5$^\circ$ and a consistent detection capability within the 4$\pi$ solid angles range. After a detailed treatment of systematic uncertainties, we find that the muon flux exceeds model expectations by approximately 40\%, with a significance of more than 2$\sigma$ when including model-related uncertainties, consistent with similar anomalies reported elsewhere. Our analysis suggests that a harder charged hadron spectrum in the earliest stages of the cascade, possibly combined with an enhancement of charged $K$ production, could account for this excess, offering specific evidence for resolving the \textit{muon puzzle}. Alternatively, assuming fixed hadronic models, our data also favor a lighter primary CR composition in the relevant energy range compared to the GSF model. We anticipate that coincident measurements between surface and underground laboratories, or combined analyses of underground and ground-based array measurements at similar cosmic-ray energy scales, can facilitate further detailed studies of primary CR composition and hadronic interactions in EAS.

\begin{acknowledgement}
We thank Zhen Cao, Huihai He, William Woodley, Ludwig Neste, and Andrey Romanov for many useful discussions on cosmic ray and extensive air shower physics. We also thank Jianqiao Wang, Yuxiang Song, Zhengchen Lian, and Xianglei Zhu for their help in studies of hadronic and nuclear interactions. We acknowledge Orrin Science Technology, Jingyifan Co., Ltd, and Donchamp Acrylic Co., Ltd, for their efforts in the engineering design and fabrication of the stainless steel and acrylic vessels. Many thanks to the CJPL administration and the Yalong River Hydropower Development Co., Ltd. for logistics and support.

\par\smallskip\noindent\textbf{Funding.} This work was supported in part by the National Natural Science Foundation of China (Grant Nos. 12127808, 12141503, 12305117, 12441513, and 12521007), the Key Laboratory of Particle and Radiation Imaging (Tsinghua University), and the State Key Research Development Program in China (Grant No. 2022YFA1604700).

\par\smallskip\noindent\textbf{Data Availability Statement.} The datasets generated during and/or analysed during the current study are available from the corresponding author on reasonable request.

\par\smallskip\noindent\textbf{Code Availability Statement.} The code/software generated during and/or analysed during the current study is available from the corresponding author on reasonable request.
\end{acknowledgement}

\bibliographystyle{spphys}
\bibliography{bibfile}

\end{document}